\documentclass[aps,pra,reprint,amsmath,amssymb,superscriptaddress,showpacs,nobibnotes]{revtex4-1}

\usepackage{graphicx}
\usepackage{bm}     
\usepackage{hyperref} 
\usepackage{xcolor}
\hypersetup{
    colorlinks,
    linkcolor={blue!80!black},
    citecolor={blue!80!black},
    urlcolor={blue!80!black}
}
\usepackage{dcolumn} 
\usepackage{dsfont}    
\usepackage{color}
\usepackage{textcomp,gensymb} 
\begin{document}

\title{Experimental characterization of quantum polarization of three-photon states }

\author{Yosep Kim}
\affiliation{Department of Physics, Pohang University of Science and Technology (POSTECH), Pohang, 37673, Korea}

\author{Gunnar Bj{\"o}rk}
\affiliation{Department of Applied Physics, Royal Institute of Technology (KTH), AlbaNova, SE-106 91 Stockholm, Sweden}

\author{Yoon-Ho Kim}
\affiliation{Department of Physics, Pohang University of Science and Technology (POSTECH), Pohang, 37673, Korea}

\date{\today}

\begin{abstract}
We experimentally investigate various quantum polarization features of three-photon quantum states, including product and entangled states with varying purity. The three-photon quantum states are categorized into six classes based on the rotation symmetry of mean, variance, and skewness of the polarization distribution. The representative three-photon quantum states in each category is prepared from double-pair emission from pulsed spontaneous parametric down-conversion and quantum interferometry.  We demonstrate that the three-photon quantum states show interesting quantum polarization properties, such as,  maximum sum-uncertainty and hidden polarizations.
\end{abstract}

\maketitle


\section{\label{sec:1}Introduction}

The polarization degree of freedom of single photons has been widely used to explore quantum phenomena, since polarization entangled photon pairs can be prepared with  high fidelity and the polarization state is easy to manipulate with linear optics. For example, polarization entangled photons have been utilized  for a variety of fundamental tests of quantum physics, including local hidden variables theories \cite{Kwiat95} and epistemic models of the wavefunction \cite{Ringbauer15}. In addition, a plethora of quantum information technologies have been experimentally implemented using photon polarization such as quantum key distribution \cite{Schmitt07}, quantum dense coding \cite{Mattle96}, quantum teleportation \cite{Kim01}, and quantum computing \cite{Lopez12}.

The polarization state of light is conventionally described through the use of Stokes parameters that can be represented as a polarization direction and a degree of polarization on  the Poincar{\'e} sphere \cite{Stokes52}. Since the Stokes parameters show only the averaged, ``classical'' features, they are not sufficient to describe the quantum polarization features fully. For instance, there exist ``classically'' unpolarized light which has non-isotropic second-order polarization, thus making the state polarized  \cite{Usachev01}. This example highlights the existence of hidden polarization features and the importance of polarization fluctuations. Up to now, various quantum states have been studied, such as, squeezed polarization states \cite{Shalm09,Klimov10} and entangled photon states \cite{Jaeger03}. In addition, efficient polarization tomography methods have been suggested \cite{Schilling10,Soderholm12,Bjork12,Bayraktar16}.

In this paper, we experimentally investigate various  quantum polarization features of three-photon quantum states, including product and entangled states with varying purity. The studied three-photon states are isomorphic to the states of a composite system consisting of three spin-1/2 particles with the bosonic characteristic, that is, the symmetric Hilbert subspace. To fully describe the properties of the states, up to the third-order Stokes parameters are necessary and sufficient, since the fourth- and higher-order polarization moments contain no additional information. The central moments of the Stokes operator is also useful to describe the polarization distribution on the Poincar{\'e} sphere with Gaussian approximation \cite{Bjork12}. The mean, variance, skewness represent the first, second, third order central moments, respectively. Through their central moments, all three-photon quantum states are categorized into six different classes according to Table~\ref{tab:table1} based on their rotation invariance (on the Poincar{\'e} sphere) of their the mean, variance and skewness.  In this work, six class-representative three-photon quantum states are experimentally prepared and measured to confirm the predicted polarization properties.

\renewcommand{\arraystretch}{1.5}
\begin{table*}
\caption{\label{tab:table1}
Classification and examples of three-photon quantum polarization based on SU(2) rotation invariance of mean $\langle \hat{S}_{\mathbf{n}} \rangle$, variance $\langle \hat{\Delta}^{2}_{\mathbf{n}} \rangle$, and skewness  $\langle \hat{\Delta}^{3}_{\mathbf{n}} \rangle$ \cite{Bjork12}. The symbol O indicates rotation invariance of the particular order of quantum polarization.}

\begin{tabular}{>{\centering}m{.1\linewidth}>{\centering}m{.1\linewidth}>{\centering}m{.1\linewidth}m{.05\linewidth}m{.4\linewidth}}

\hline\hline
 \multicolumn{3}{c}{Rotation invariance\footnote{For three-photon polarization states, six classes are physically possible among eight possibilities.}}&\\ \cline{1-3}

 $\langle \hat{S}_{\mathbf{n}} \rangle$&$\langle \hat{\Delta}^{2}_{\mathbf{n}} \rangle$&$\langle \hat{\Delta}^{3}_{\mathbf{n}} \rangle$&&\multicolumn{1}{c}{\ \ Representative state in each class\footnote{Fock states in horizontal and vertical polarization mode.}}\\ \hline

O&O&O&&$\ \ \ \ \hat{\mathbb{I}}/4$\\
O&O&X&&$\ \ \ \ \frac{1}{3}\!\left |3,0\rangle\langle 3,0 \right |+\frac{1}{2}\!\left | 1,2\rangle\langle 1,2 \right |+\frac{1}{6}\!\left | 0,3\rangle\langle 0,3 \right |$\\
O&X&O&&$\ \ \ \ \frac{1}{2}\!\left ( \left | 3,0\rangle\langle 3,0 \right |+\left | 0,3\rangle\langle 0,3 \right |\right )$\\
O&X&X&&$\ \ \ \ \frac{1}{\sqrt{2}}\!\left ( | 3,0\rangle-i| 0,3\rangle\right )$\\
X&O&X&&$\ \ \ \ \frac{19}{36}\!\left |3,0\rangle\langle 3,0 \right |+\frac{15}{36}\!\left | 1,2\rangle\langle 1,2 \right |+\frac{1}{18}\!\left | 0,3\rangle\langle 0,3 \right |$\\
X&X&X&&$\ \ \ \ | 3,0\rangle$\\
\hline\hline
\end{tabular}
\end{table*}


\section{\label{sec:2}Theory}
The Stokes parameters consist of the total intensity $S_{0}$ and the three elements of Stokes vector $\vec{S}=(S_{1},S_{2},S_{3})$ which represent complementary polarization directions on the Poincar{\'e} (or Bloch) sphere. The values of $(S_{1},S_{2},S_{3})$  are obtained from intensity differences between orthogonal polarizations: diagonal/anti-diagonal, right/left circular, and horizontal/vertical polarizations, respectively. By substituting the bosonic number operator for intensity, the Stokes operators can be well defined and they give quantized values of the Stokes parameters. The operators are expressed as \cite{Usachev01}
\begin{equation}
\begin{split}\label{eq:01}
&\hat{S}_{0}=\hat{a}^{\dagger}_{H}\hat{a}_{H}+\hat{a}^{\dagger}_{V}\hat{a}_{V}, \ \ \ \ \ \ \hat{S}_{1}=\hat{a}_{H}\hat{a}^{\dagger}_{V}+\hat{a}^{\dagger}_{H}\hat{a}_{V}, \\
&\hat{S}_{2}=i(\hat{a}_{H}\hat{a}^{\dagger}_{V}-\hat{a}^{\dagger}_{H}\hat{a}_{V}),\ \ \ \hat{S}_{3}=\hat{a}^{\dagger}_{H}\hat{a}_{H}-\hat{a}^{\dagger}_{V}\hat{a}_{V}
\end{split}
\end{equation}
where $\hat{a}_{H}$ $(\hat{a}_{V})$ is the annihilation operator for the horizontal (vertical) polarization mode. Their commutation relationships can be derived from the bosonic commutation relations between annihilation operators.
\begin{subequations}
\begin{eqnarray}
&[&\hat{S}_{0},\hat{S}_{j}]=0,\label{eq:02a}\\
&[&\hat{S}_{j},\hat{S}_{k}]=i2\epsilon_{jkl}\hat{S}_{l}, \ \ \ \ \ \ j,k,l \in \{1,2,3\} \label{eq:02b}
\end{eqnarray}
\end{subequations}
where $\epsilon_{jkl}$ is the Levi-Civita tensor symbol.

The physical properties of the Stokes operators are implied by the commutation relations. As indicated by Eq.~(\ref{eq:02a}), the commutation between the total photon number operator $\hat{S}_{0}$ and all other Stokes operators $\hat{S}_{j}$, indicates that the Stokes parameters and photon number can be measured independently without mutual disturbance. This allows measurements of the Stokes parameters for specific photon-number states with photon-number resolving detectors. This also implies that any moment of the Stokes operators $\hat{S}_{1}$, $\hat{S}_{2}$, and $\hat{S}_{3}$ can be measured in a similar manner.

Moreover, Eq.~(\ref{eq:02b}) indicates that some Stokes parameters must be uncertain, leading to the following inequalities,
\begin{subequations}
\begin{eqnarray}
&\sqrt{\langle\hat{\Delta}_{j}^2 \rangle}\sqrt{\langle \hat{\Delta}_{k}^2\rangle}\geq\left | \epsilon_{jkl}\langle \hat{S}_{l} \rangle \right |,\ \ \ j,k,l \in \{1,2,3\}\label{eq:03a}\\
&2\langle\hat{S}_{0}\rangle\leq \langle\hat{\Delta}_{1}^2\rangle+\langle\hat{\Delta}_{2}^2\rangle+\langle\hat{\Delta}_{3}^2\rangle \leq \langle \hat{S}_{0} \rangle (\langle \hat{S}_{0}\rangle+2)\label{eq:03b}
\end{eqnarray}
\end{subequations}
where $\Delta_{j} \equiv \hat{S}_{j}-\langle \hat{S}_{j} \rangle$ is the central-moment of the Stokes operator $j$. The existence of a photonic quantum state with hidden polarization, i.e., high-order central-moments may have non-zero values even though the state may be first-order unpolarized, $\langle \hat{\vec{S}} \rangle=0$, is implied in Eq.~(\ref{eq:03a}).    Note that, Eq.~(\ref{eq:03b}) gives boundaries of the sum of the second-order central moments. For three-photon quantum states, the sum is bounded between 6 and 15.

The notion of Stokes operators can be generalized by defining
\begin{equation}
\hat{S}_{\mathbf{n}} = (\hat{S}_{1},\hat{S}_{2},\hat{S}_{3}) \cdot \mathbf{n},
\end{equation}
where $\mathbf{n}$ is a unit vector on the Poincar\'{e}  sphere. The operator $\hat{S}_{\mathbf{n}}$ assesses the polarization state of the photonic quantum state in the direction $\mathbf{n}$. It follows trivially that one can define polarization of order $m$ in direction $\mathbf{n}$ as $\langle \Delta_{\mathbf{n}}^m  \rangle$ where $\Delta_{\mathbf{n}} \equiv \hat{S}_{\mathbf{n}}-\langle \hat{S}_{\mathbf{n}} \rangle$.


\section{\label{sec:3}Experiment}

\begin{figure}[tp]
\centering
\includegraphics[width=3.4in]{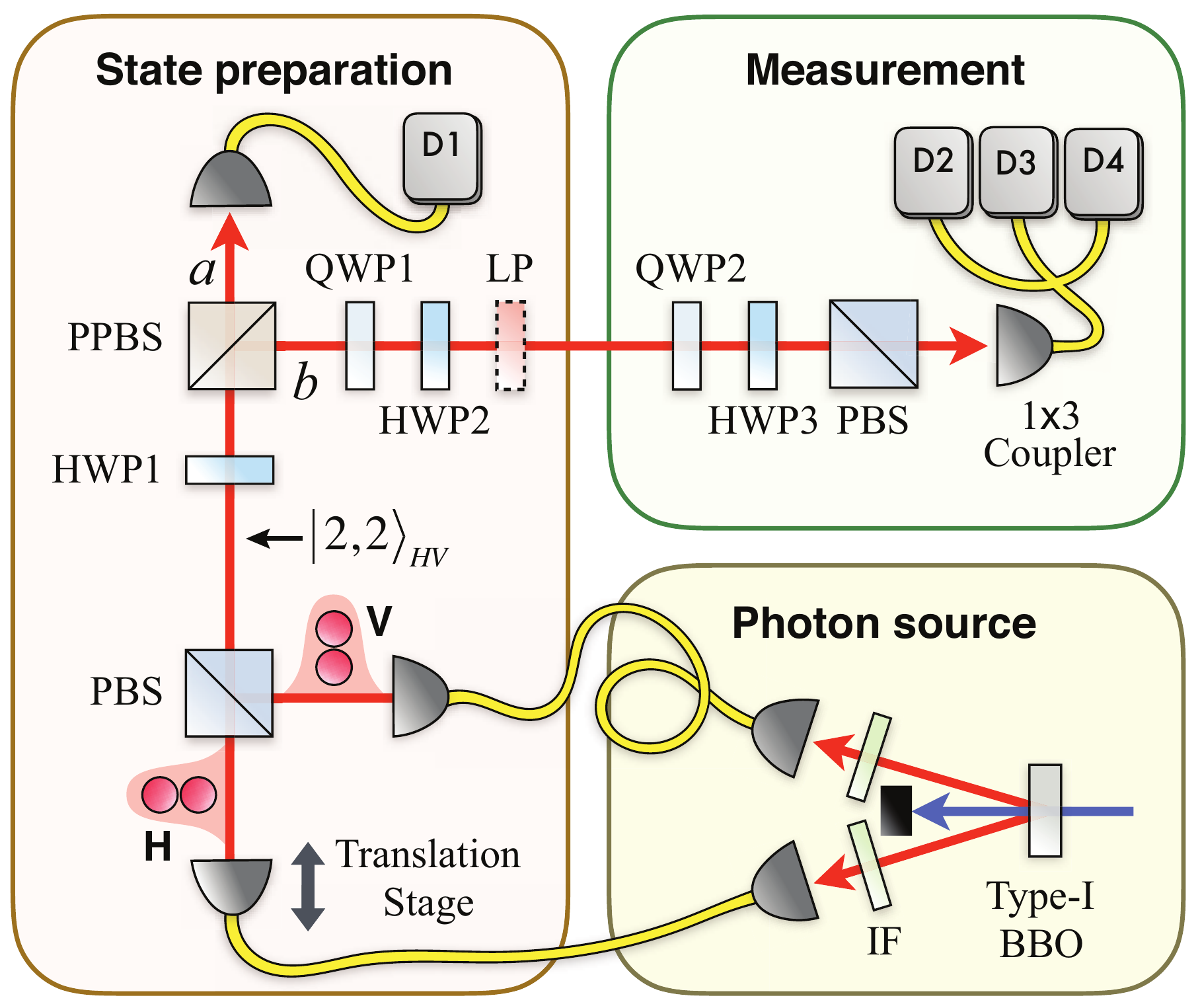}
\caption{Experimental scheme for generation and measurement of three-photon states. Two pairs of SPDC photons are sent to an interferometer through single mode fibers (SMF) after passing 3 nm bandwidth interference filters (IF). In the state preparation interferometer, the target states are prepared with a polarizing beam splitter (PBS), a partially-polarizing beam splitter (PPBS), half- and quarter-wave plates (HWP, QWP), and a linear polarizer (LP). Conditioned on the detection of a single photon at detector D1, three photons are prepared in mode $b$ in a particular quantum state set by QWP1, HWP2, and LP. Four-fold coincidence measurements with detectors D1, D2, D3, and D4 for sixteen polarization projection measurement allow quantum state tomography for the heralded three-photon states.
}\label{fig:fig1}
\end{figure}

To confirm the quantum polarization features of three-photon quantum states experimentally, six representative three-photon quantum states are prepared, as shown in Table~\ref{tab:table1}. The three-photon states are generated from the double pair emission of femtosecond-pulse-pumped spontaneous parametric down-conversion (SPDC), see Fig.~\ref{fig:fig1} \cite{Kim09,Kim11}. The pump pulse is derived from a frequency-doubled mode-locked Ti:Sapphire laser and has a central wavelength of 390 nm, a pulse duration of 140 fs, and a repetition rate of 80 MHz. The SPDC photons, generated from a type-I BBO crystal, have a central wavelength of 780 nm and propagates non-collinearly with the pump laser.

\begin{figure}[tp]
\centering
\includegraphics[width=3.4in]{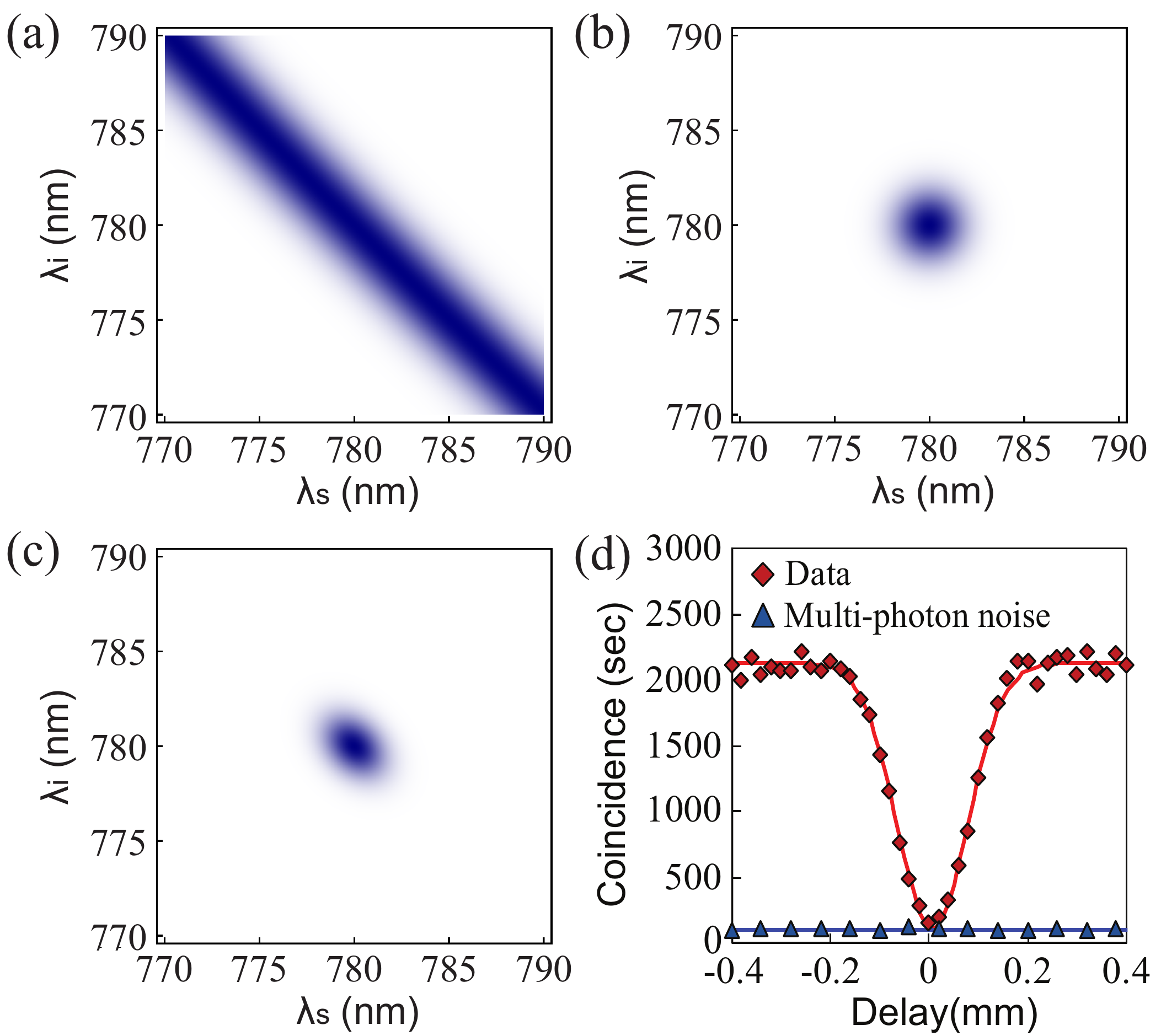}
\caption{The simulated joint spectra of (a) SPDC photons, (b) interference filters with 3 nm FWHM bandwidth,  and (c) SPDC photons filtered with the interference filters.  The simulated joint spectrum in (c) shows that the spectral correlation between the photon pair is well eliminated by the interference filters. (d) The Shih-Alley/Hong-Ou-Mandel interference dip has the visibility of 95.0\% (99.6\% after multi-photon noise subtraction) at 260 mW pump power, a clear experimental indication that spectral distinguishability between the SPDC photons have been  well eliminated  by the interference filters. The red solid line is the Gaussian fit to the data. The blue solid line is the linear fit to the calculated multi-photon noise.
}\label{fig:fig2}
\end{figure}

To prepare a heralded three-photon quantum state via interference using the scheme in Fig.~\ref{fig:fig1}, it is essential that the inherent frequency correlation between the SPDC photon pair be eliminated \cite{Grice01}. In our work, we ensure that this condition is satisfied by using a 0.6 mm thick type-I BBO crystal, generating broadband SPDC photons, and by filtering the SPDC photons with interference filters (IF) with full width at half maximum (FWHM) of 3 nm. The calculated spectral properties of the unfiltered SPDC photons are shown in Fig.~\ref{fig:fig2}(a). It is clear that the SPDC photons from a 0.6 mm thick type-I BBO has a very broadband emission and show a very strong spectral correlation between the pair. When the SPDC photons are filtered with the 3 nm interference filters, see Fig.~\ref{fig:fig2}(b), the resulting SPDC photons have almost no spectral correlations, see Fig.~\ref{fig:fig2}(c).

The non-collinear, double pair SPDC photons are combined into a single spatial mode by a PBS through single-mode optical fibers for spatial mode cleaning, see Fig.~\ref{fig:fig1}. Initially, all four photons are horizontally polarized but the use of a fiber polarization controller allows us to combine all four photons to a single spatial mode without loss. As all four photons must also be indistinguishable temporally, they all need to arrive at PBS simultaneously. This has been achieved by observing the Shih-Alley/Hong-Ou-Mandel dip between the SPDC photon at the PBS \cite{Shih86,Hong87}. To observe the two-photon interference dip, the angles of HWP1 and HWP3 are set, respectively, at 22.5$\degree$ and 45$\degree$. All other wave plates, HWP2, QWP1 and QWP2, are set at 0$\degree$ and LP is removed. the coincidence between the detectors D1 and D2 are measured by moving the translation stage on horizontal input mode in Fig.~\ref{fig:fig1}. The experimental result shown in  Fig.~\ref{fig:fig2}(d) exhibits the dip visibility of 95.0\% at 260 mW pump power (99.6\% after multi-photon noise subtraction). The translation stage is then set so that the SPDC photons are arriving at the PBS simultaneously. Then, the four-photon quantum state, resulting from the double-pair emission of the SPDC, after the PBS is written as  $|2,2\rangle_{H,V}$.

We now describe the scheme for heralding a three-photon quantum state in mode $b$ by detecting a single photon at D1. The initial four-photon state  $|2,2\rangle_{H,V}$ passes through HWP1 and the HWP1 angle (0$\degree$ or 22.5$\degree$) is set differently for preparing different  three-photon states. Specifically, HWP1 is set at 0$\degree$ to prepare  $|1,2\rangle$ and $|2,1\rangle$  and set at 22.5$\degree$ to prepare $\frac{1}{\sqrt{2}}\!(|3,0\rangle-i|0,3\rangle), |3,0\rangle$, and $|0,3\rangle$. After HWP1, the state becomes
\begin{subequations}
\begin{eqnarray}
&|&H\!W\!\!P1\rangle_{0\degree} \ \ = \frac{1}{2}{a^{\dagger\ 2}_H}{a^{\dagger\ 2}_V}|0\rangle,\label{eq:04a}\\
&|&H\!W\!\!P1\rangle_{22.5\degree} \!=\!\!\left (\frac{1}{8}{a^{\dagger\ 4}_H}-\frac{1}{4}{a^{\dagger\ 2}_H}{a^{\dagger\ 2}_V}+\frac{1}{8}{a^{\dagger\ 4}_V}\right )\!|0\rangle.\label{eq:04b}
\end{eqnarray}
\end{subequations}
The subscripts 0$\degree$ and 22.5$\degree$ indicate the angles of  HWP1. The photons then impinge on the partially-polarizing beam splitter (PPBS) designed for unity reflection for vertical polarization and  1/3 partial reflection for horizontal polarization. Considering the case when one photon is transmitted and found in mode $a$ and three photons are reflected by the PPBS and found in mode $b$, the reflected three-photon state heralded by the presence of a single-photon in mode $a$ is given by
\begin{subequations}
\begin{eqnarray}
&|&P\!P\!B\!S\rangle_{0\degree}^b \ \ = \frac{1}{\sqrt{2}}{a^{\dagger}_H}{a^{\dagger\ 2}_V}|0\rangle,\label{eq:05a}\\
&|&P\!P\!B\!S\rangle_{22.5\degree}^b \!=\!\!\left ( \frac{1}{9\sqrt2}e^{2i\phi}{a^{\dagger\ 3}_H}-\frac{1}{3\sqrt{2}}{a^{\dagger}_H}{a^{\dagger\ 2}_V}\right )\!|0\rangle,
\label{eq:05b}
\end{eqnarray}
\end{subequations}
where the phase $\phi$ comes from the relative phase-difference between the two orthogonal polarizations when they are reflected at the PPBS.

\begin{figure*}[tp]
\centering
\includegraphics[width=6.5in]{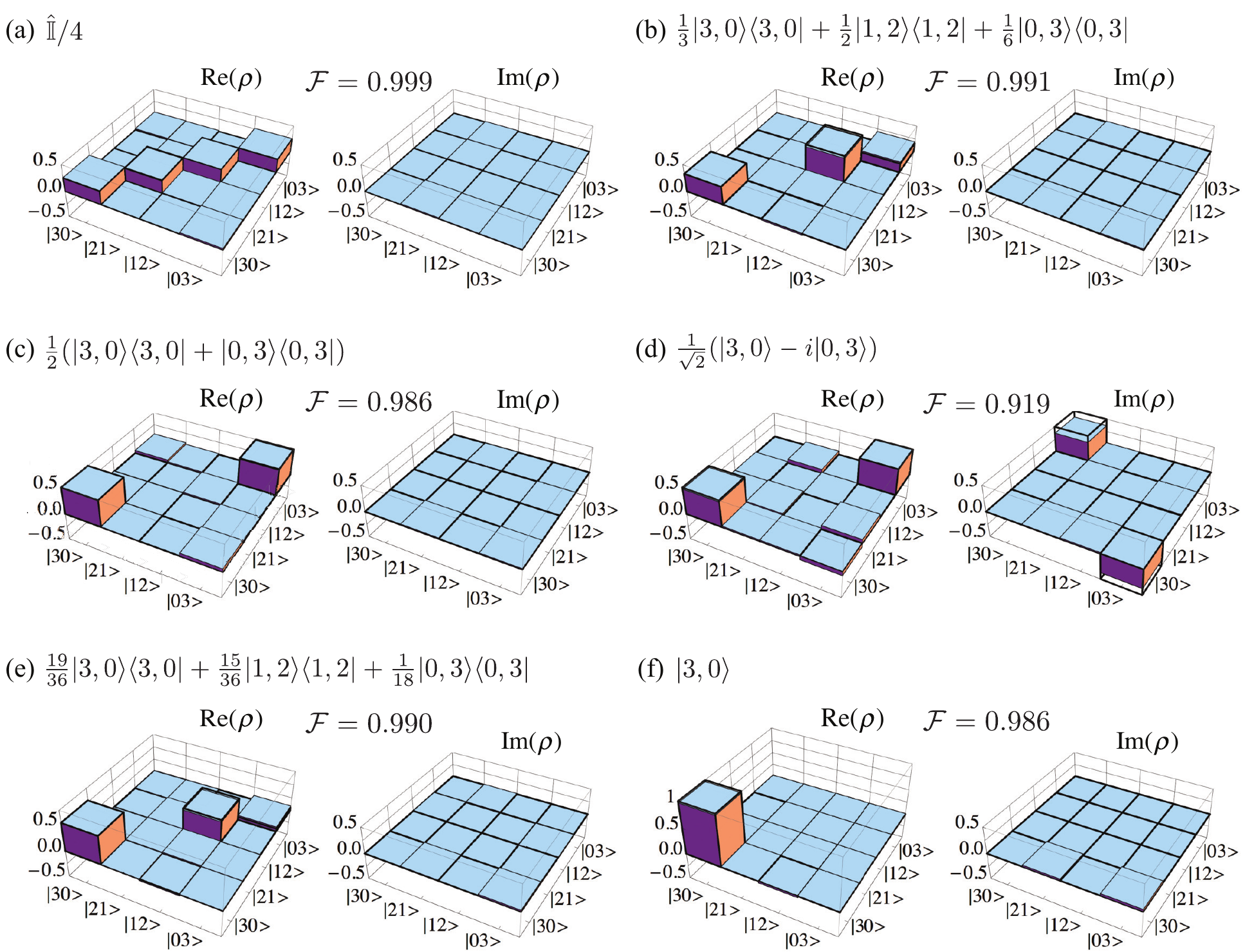}
\caption{Density matrices of experimentally prepared heralded three-photon states. Bold  lines in the density matrices indicate ideal target density matrices. The fidelity is calculated between the ideal target matrix and the experimentally generated density matrix.
}\label{fig:fig3}
\end{figure*}

The transmitted photon in mode $a$ is used for heralding of the other three photons by a ``click'' at detector D1.  Then, after the photons pass through QWP1 and HWP2, either both set at 0$\degree$ for Eq.~(\ref{eq:06a})
or QWP1 at $45\degree$ and HWP2 at $\phi/4$ for Eq.~(\ref{eq:06b}), the heralded three-photon state becomes,
\begin{subequations}
\begin{eqnarray}
&|&H\!W\!\!P2\rangle_{0\degree} \ \ =\frac{1}{\sqrt{2}}{a^{\dagger}_H}{a^{\dagger\ 2}_V}|0\rangle\label{eq:06a},\\
&|&H\!W\!\!P2\rangle_{22.5\degree} \!=\frac{1}{2\sqrt{3}}\left ({a^{\dagger\ 3}_H}-i{a^{\dagger\ 3}_V}\right )\!|0\rangle.\label{eq:06b}
\end{eqnarray}
\end{subequations}
Here, we see that the heralded three photon states $|1,2\rangle$ and $\frac{1}{\sqrt{2}}\!(|3,0\rangle-i|0,3\rangle)$ have been prepared. The state $|2,1\rangle$ can be prepared from $|1,2\rangle$ with the help of HWP2 set at 45$\degree$.  Also, the states $|3,0\rangle$ and $|0,3\rangle$  can be post-selected from the entangled state or the N00N state $\frac{1}{\sqrt{2}}\!(|3,0\rangle-i|0,3\rangle)$ with a linear polarizer (LP).

We first measure the the value of $\phi$ experimentally  by using the $|1,1\rangle_{H,V}$ component of the SPDC leading to the coincidence event between detectors D1 and D2. First, HWP1 and QWP1 are set at 15$\degree$ and $45\degree$, respectively. The photon in mode $b$ is measured on the projection basis $\frac{1}{\sqrt{2}}(|H\rangle - |V\rangle)$. Then, as a function of the HWP2 angle $\theta$, the coincidence count between detectors D1 and D2 will be proportional to $\sin^{2}\{\frac{1}{2}(\phi-4\theta)\}$. Thus, by measuring the angle $\theta$ at which the coincidence count is minimized, it is possible to determine the phase $\phi$. Experimentally, we find that $\phi=-85.7\degree$.

In our scheme, the double-pair event of SPDC contributes to the heralded three-photon state but triple-pair or higher order SPDC events lead to multi-photon noise as they can also trigger the four-photon coincidence circuit. As $N$-pair events of SPDC with $N \geq 3$ increase with the double-pair event of SPDC, one needs to consider the trade-off between the detection rate and the multi-photon noise contribution to the data. In our setup, we find that 260 mW pump power results in 2.5\%, 0.06\%, and 0.002\% emission probabilities of a single, double, triplet pairs, respectively. In case of the states $|3,0\rangle$ and $|0,3\rangle$, on the other hand, the four-fold coincidence probability is half of the state $\frac{1}{\sqrt{2}}\!(|3,0\rangle-i|0,3\rangle)$. Thus, we are able to use twice high pump power in this case to reduce the measurement time.

Finally, the heralded three-photon state is characterized by performing quantum state tomography using  16 projection measurements set by QWP2, HWP3, and PBS in  Fig. \ref{fig:fig1} and maximum likelihood estimation \cite{Israel12}. The experimentally obtained density matrices for the representative three-photon states listed in Table I are shown in Fig.~\ref{fig:fig3}. Mixed states are obtained by incoherently adding pure states with the proper ratio. The three-photon entangled state $\frac{1}{\sqrt{2}}\!(|3,0\rangle-i|0,3\rangle)$ has the lowest   fidelity because the triple-pair contribution from pulsed SPDC to the non-interfering background is more noticeable \cite{Kim09,Kim11}.

\section{\label{sec:4}Analysis}

In this section, we analyze the experimentally generated three-photon polarization states with the central-moment description of quantum polarization up to the third order. (The fourth and higher order Stokes moments can be expressed as function of the lower order moments.) The central-moments, plotted in the corresponding direction on the Poincar{\'e} sphere, shows quantum polarization features such as rotation invariants and the polarization uncertainty distribution. For instance, classical unpolarized light is SU(2) rotation invariant, that is, it is rotationally invariant in all directions on the sphere. If the first $\mathit{m}$ moments are rotationally invariant, the state is defined to be $\mathit{m}$-th order unpolarized \cite{Bjork12}. As for the uncertainty relation regarding quantum polarization in Eq.~(\ref{eq:03b}), all three-photon states have a common restriction that the sum of the variances along three orthogonal polarizations on the Poincar{\'e} sphere (indicating three complementary polarization bases) is between 6, from $2\langle \hat{S_0}\rangle$, and 15, from $\langle \hat{S_0}\rangle (\langle \hat{S_0}\rangle+2))$ because $\langle \hat{S}_{0}\rangle= 3$.

\begin{figure*}[tp]
\centering
\includegraphics[width=6.8in]{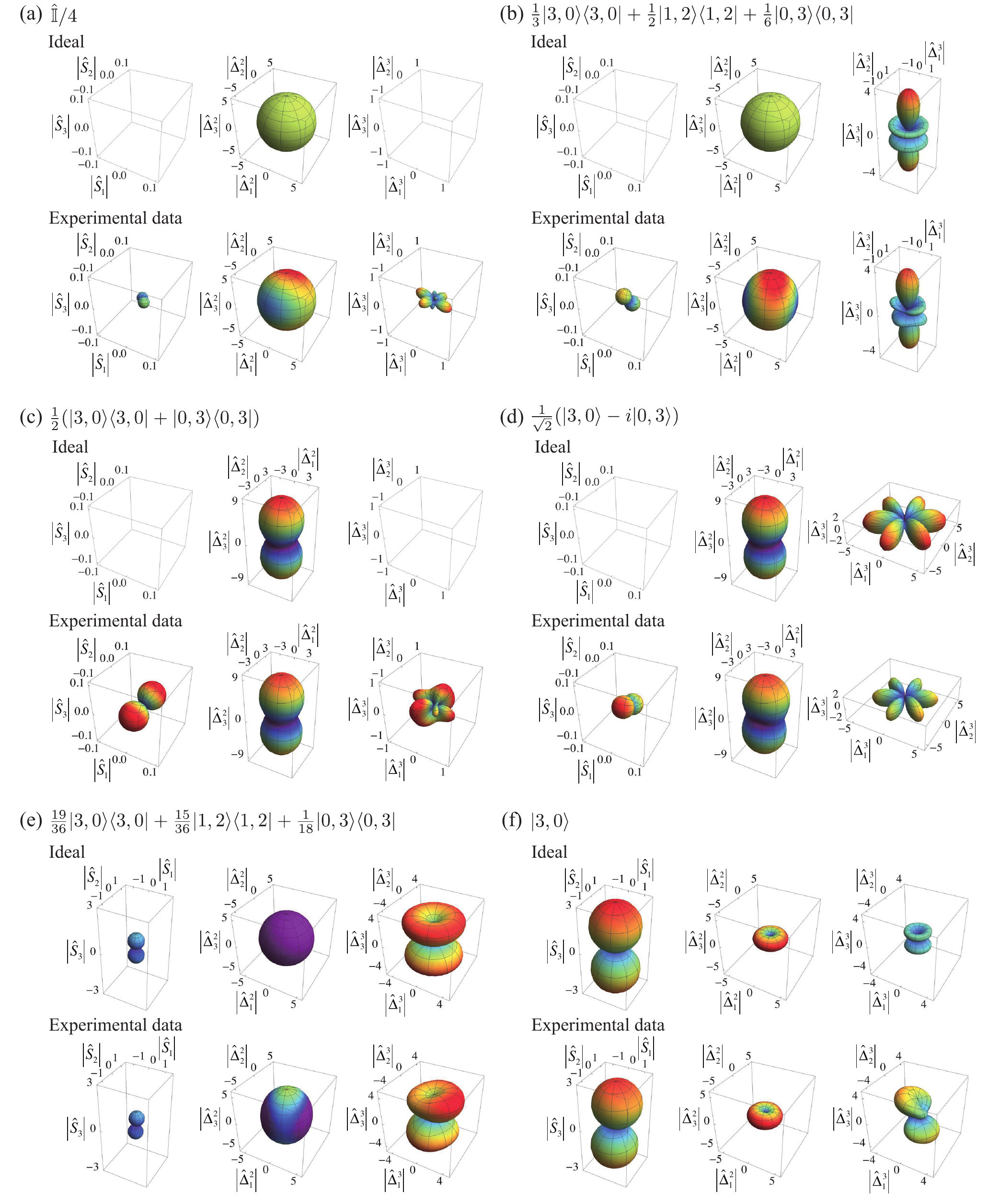}
\caption{Central-moment quantum polarization descriptions of assorted three-photon states. The rainbow-colored gradient represents the distance from the origin of the coordinate system. In general, the red color  illustrates the farther distance from the origin.  Depicted in each panel are, left-to-right, the mean value $\langle |\hat{S}_{\mathbf{n}}|\rangle$, the variance $\langle \hat{\Delta}^{2}_{\mathbf{n}} \rangle$, and the absolute value of the skewness $\langle |\hat{\Delta}^{3}_{\mathbf{n}}| \rangle$). Note that, for odd polarization orders $m$, $\langle \Delta_{\mathbf{n}}^m \rangle = -\langle \Delta_{\mathbf{-n}}^m \rangle$ and, thus, the illustrated quantity is the absolute value of the moment plotted as a function of the direction $\mathbf{n}$ on the Poincar\'{e} sphere. }
\label{fig:fig4}
\end{figure*}

The identity density matrix is invariant of any rotation transformation since $\hat{\mathbb{I}}/4=\hat{U}^{\dagger}_{\mathbf{n}}(\hat{\mathbb{I}}/4)\hat{U}_{\mathbf{n}}$ where $\hat{U}_{\mathbf{n}}$ is any SU(2) rotation. So, this state always gives an isotropic expectation value for quantum polarization in any basis. That is, every order of its quantum polarization is rotationally invariant and the state is hence unpolarized to every order. Since the state is first-order unpolarized, the second-order central-moment or the variance, is given by the mean square of the Stokes operator. As the eigenvalues of $\hat{S}_{3}$ for the states $|3,0\rangle, |2,1\rangle, |1,2\rangle$, and $|3,0\rangle$ are $3, 1, -1,$ and $-3$, respectively, the value $\langle \hat{\Delta}^{2}_{3}\rangle$ for the state $\hat{\mathbb{I}}/4$ is $(9+1+1+9)/4 =5$. This state has the isotropic variance, so the sum of variance is 15, making it the maximum sum-uncertainty state. As expected, in Fig.~\ref{fig:fig4} (a), the experimental result confirms the theoretical results with only small errors. Note that for the odd polarization orders, the illustrated quantity is the absolute value of the moment plotted as a function of the direction $\mathbf{n}$ on the Poincar\'{e} sphere. For odd polarization orders $m$ it holds that $\langle \Delta_{\mathbf{n}}^m \rangle = -\langle \Delta_{\mathbf{-n}}^m \rangle$.

Any second-order unpolarized three-photon quantum state must be a mixed state, see the proof in \cite{Bjork12}. Among the mixed states, the state $\frac{1}{3}\!\left |3,0\rangle\langle 3,0 \right |+\frac{1}{2}\!\left | 1,2\rangle\langle 1,2 \right |+\frac{1}{6}\!\left | 0,3\rangle\langle 0,3 \right |$ has the maximum sum-uncertainty. This state is isotropic up to the second-order and is polarized in the third-order, as shown theoretically and experimentally in Fig.~\ref{fig:fig4}(b).

The states $|3,0\rangle$ and $|0,3\rangle$ have rotation symmetry about the $\hat{S}_{3}$ axis. The symmetry is preserved upon mixing (i.e., incoherent addition of quantum states) and they have vectors in opposite directions from each other on the Poincar{\'e} sphere. Therefore, the mixed state $\frac{1}{2}\! \left ( \left | 3,0\rangle\langle 3,0 \right |+\left | 0,3\rangle\langle 0,3 \right |\right )$ has vanishing all odd-order central moments as shown in Fig.~\ref{fig:fig4}(c).  The second-order central moment is however maximized along the $\hat{S}_{3}$ axis (with the value of 9) and minimized on the $\hat{S}_{1}$-$\hat{S}_{2}$ plane (with the value of 3). The state $\frac{1}{2}\! \left ( \left | 3,0\rangle\langle 3,0 \right |+\left | 0,3\rangle\langle 0,3 \right |\right )$ therefore is a three-photon maximum sum-uncertainty state, having the quantum polarization properties of first-order hidden polarization, polarized in the second-order, and with no third-order polarization.

Let us now consider the case of the entangled state $\frac{1}{\sqrt{2}}\!\left ( | 3,0\rangle-i| 0,3\rangle\right )$.  It is clear that some quantum polarization features of the entangled state would be similar to the mixed state  $\frac{1}{2}\! \left ( \left | 3,0\rangle\langle 3,0 \right |+\left | 0,3\rangle\langle 0,3 \right |\right )$ as both contain the same basis states. This is reflected in the first-order and second-order quantum polarization properties. The key difference between the two states are inherent coherence and is reflected on the third-order quantum polarization, producing skewness in three directions as shown in Fig.~\ref{fig:fig4}(d). The three-photon N00N state is well-known to exhibit $N$ times phase sensitivity compared to a classical state and this feature is illustrated in the third-order quantum polarization, showing three oscillations during the 2$\pi$ phase change on the $\hat{S}_{1}$-$\hat{S}_{2}$ plane. Note that the figure shows the absolute value of central moment. In the experiment, the state is not quite ideal, see the density matrix in Fig.~\ref{fig:fig3}(d), so the skewness is somewhat reduced (resulting in reduced N00N state interference visibility) compared to the theoretical one. Note that the variances  $\langle \Delta^{2}_{1} \rangle, \langle \Delta^{2}_{2} \rangle$, and $ \langle \Delta^{2}_{3} \rangle$ are calculated to be $3, 3,$ and $9$, respectively. Thus, the state is also a  maximum sum-uncertainty state. The state has first-order hidden polarization but is polarized to second- and third-order.

We now consider the state $\frac{19}{36}\!\left |3,0\rangle\langle 3,0 \right |+\frac{15}{36}\!\left | 1,2\rangle\langle 1,2 \right |+\frac{1}{18}\!\left | 0,3\rangle\langle 0,3 \right |$ which is first-order and third-order polarized, but with isotropic second-order central moment, see Table I. The theoretical and experimental results shown in Fig.~\ref{fig:fig4}(e) illustrate this feature of the state. Note that, since the state is a mixture of horizontal and vertical basis eigenstates, it has rotational symmetry about the $\hat{S}_{3}$ axis. But, it is not a maximal sum-uncertainty state.

Finally, consider the state  $| 3,0\rangle$ which is clearly first-order, second-order, and third-order polarized as all three photons are horizontally polarized and this is shown by the anisotropic features in all orders of central moments in Fig.~\ref{fig:fig4}(f). Since the state is an eigenstate of $\hat{S}_{3}$,  $\langle\hat{\Delta}_{3}^{m}\rangle$ vanishes for all $m$. Moreover,  both $\langle\hat{\Delta}_{1}^{m}\rangle$ and $\langle\hat{\Delta}_{2}^{m}\rangle$ vanish for odd $m$. Thus the odd polarization moments all vanish on the $\hat{S}_1$-$\hat{S}_2$ plane. Note that the state satisfies (\ref{eq:03a}) with equality on the $\hat{S}_{1}$-$\hat{S}_{2}$ plane.

The advantages of using the quantum polarization description of the multi-photon state can be summarized as follows. As evidenced in Fig.~\ref{fig:fig4}, the quantum polarization description allows one to visually identify for which applications the quantum state is best suited. For instance, the mixed state $\frac{1}{3}\!\left |3,0\rangle\langle 3,0 \right |+\frac{1}{2}\!\left | 1,2\rangle\langle 1,2 \right |+\frac{1}{6}\!\left | 0,3\rangle\langle 0,3 \right |$ shown in Fig.~\ref{fig:fig4}(b) can be useful for polarization interferometry involving three-photon correlation measurement. However, the mixed state $\frac{1}{2}\! \left ( \left | 3,0\rangle\langle 3,0 \right |+\left | 0,3\rangle\langle 0,3 \right |\right )$ shown in Fig.~\ref{fig:fig4}(c) is better suited for two-photon correlation interferometry due to the anisotropy in the second-order quantum polarization. Also, the N00N state offers the best phase sensitivity to SU(2) rotations, as evidenced in the third-order quantum polarization behavior shown in Fig.~\ref{fig:fig4}(d).  Note also from Fig.~\ref{fig:fig4}(d) that  the N00N state offers three-fold improvement of phase sensitivity over the classical behavior as well as the possibility to get unity interference visibility. Such information is not at all evident from the density matrix description of the quantum states shown in Fig.~\ref{fig:fig3}, although, for a two-mode state with $N$-photons, or equivalent, a state of composite system for $N$ spin-$1/2$ particles, the two figures contain mathematically equivalent and interconvertible information. Additionally, deviations of the experimental quantum states from their ideal target states are more easily identified in Fig.~\ref{fig:fig4} than in Fig.~\ref{fig:fig3} as it is difficult to deduce such information from a visual inspection of the density matrices in Fig.~\ref{fig:fig3}.


\section{\label{sec:5}Conclusion}

We have experimentally studied diverse quantum polarization features of different three-photon states, selected to represent the six possible three-photon polarization classes.  The states have interesting characteristics such as perfect polarization, absence of polarization, hidden polarization and maximum sum-uncertainty. Our classification and experimental results for three-photon states can be applied as well to describe the spin features of composite systems consisting of three spin-1/2 particles with the bosonic characteristic. In addition, we have shown that  subtle quantum polarization features are more sensitive to state imperfections than those of state density matrices. Our results hint that the central moment description can be used to describe the quality of a multi-photon polarization state with better sensitivity, in particular in the cases where higher order polarization features are important.

\

\section*{Acknowledgments}
This work was supported by the National Research Foundation of Korea (Grant Nos. 2016R1A2A1A05005202, 2016R1A4A1008978, and NRF-2012K1A2B4A01033433). Y.K. acknowledges support from the Global Ph.D. Fellowship by the National Research Foundation of Korea (Grant No. 2015H1A2A1033028). G.B. acknowledges financial support by Swedish Foundation for International Cooperation in Research and Higher Education (STINT), the Swedish Research Council (VR) through its Linnaeus Center of Excellence ADOPT and Contract No. 621-2014-5410.


\pagebreak
\end{document}